\documentclass[12pt]{article}
\usepackage{graphicx} 
\usepackage{amsmath} 
\usepackage{cite}
\usepackage{wrapfig}
\usepackage{doublespace}
	\newtheorem{theorem}{Theorem}

	\newtheorem{lemma}{Lemma}

\newenvironment{example}{\par\begin{quotation}\small\noindent{\bf Example:\ }}{\end{quotation}\par}

\newenvironment{Proof}[1]{\medskip\par\noindent 
{\bf Proof:\, #1}\,}{{\mbox{\,$\bullet$}\par}}

	\newcommand{\thmref}[1]{(\ref{thm:#1})}
	\newcommand{\Thmref}[1]{Theorem~\ref{thm:#1}}
	\newcommand{\thmlabel}[1]{\label{thm:#1}}

	\newcommand{\twodef}[4]{\left\{\begin{array}{ll} 
		\displaystyle {#1} & {#2} \\
		\displaystyle {#3} & {#4}
		\end{array}\right.}
	\newcommand{\threedef}[6]{\left\{\begin{array}{ll} 
		\displaystyle {#1} & {#2} \\
		\displaystyle {#3} & {#4}\\
		\displaystyle {#5} & {#6}
		\end{array}\right.}

\newcommand{\lemlabel}[1]{\label{lem:#1}}

\newcommand{\bG}{{\bar{G}}}

\newcommand{\blankout}[1]{}

\newcommand{\be}{\begin{equation}}
\newcommand{\ee}{\end{equation}}
\newcommand{\equat}[1]{equation (\ref{eq:#1})}
\newcommand{\Equat}[1]{Equation (\ref{eq:#1})}

\newcommand{\xv}{{\bf x}}
\newcommand{\sv}{{\bf s}}
\newcommand{\vv}{{\bf v}}

\newcommand{\gv}{{\mathbf{g}}}

\newcommand{\tv}{{\bf t}}

\newcommand{\uv}{{\bf u}}

\newcommand{\Hup}{{H^{\uparrow}}}
\newcommand{\tvv}{\vec{\tv}}
\newcommand{\tvec}{\vec{t}}
\newcommand{\Tvec}{{\vec{T}}}
\newcommand{\svec}{\vec{s}}
\newcommand{\Svec}{\vec{S}}
\newcommand{\svv}{\vec{\sv}}
\newcommand{\Sv}{\vec{S}}
\newcommand{\Svv}{\vec{\Smat}}
\newcommand{\Tvv}{\vec{\Tmat}}

\newcommand{\Xvv}{\vec{\Xmat}}

\newcommand{\bzero}{{\bf 0}}
\newcommand{\binfty}{{\boldsymbol{\infty}}}

\newcommand{\Dmat}{{\bf D}}

\newcommand{\Tmat}{{\bf T}}
\newcommand{\Smat}{{\bf S}}
\newcommand{\Xmat}{{\bf X}}

\newcommand{\na}{{\Omega}}

\title{Timing Channels with Multiple Identical Quanta}
\author{Christopher Rose, Saira Mian \& Ruochen Song}
\begin{document}
\pagestyle{empty}
\maketitle
\begin{center}
{\bf Abstract}
\end{center}
{\small \em We consider mutual information between release times and capture times for a set
of $M$ identical quanta traveling independently from a source to a target.  The
quanta are immediately captured upon arrival, first-passage times are assumed
independent and identically distributed and the quantum emission times are
constrained by a deadline.  The primary application area is intended to be
inter/intracellular molecular signaling in biological systems whereby an
organelle, cell or group of cells must deliver some message (such as
transcription or developmental instructions) over distance with reasonable
certainty to another organelles, cells or group of cells.  However, the model
can also be applied to communications systems wherein indistinguishable signals
have random transit latencies.}


\pagenumbering{arabic}
\section{Introduction}
Biological systems are networks of intercommunicating elements at whatever level
one cares to consider -- (macro)molecules, cells, tissues, organisms,
populations, microbiomes, ecosystems, and so on. It is no wonder therefore that
communication theorists have plied their trade heavily in this scientific domain
(for a recent review, see \cite{todd_ITspecial10}). Biological systems offer a
dizzying array of processes and phenomena through which the same and different
tasks, communication or otherwise, might be accomplished (see, for example,
\cite{Joussineau_filopod,GorYan06,GurBar08,WanVer10,purcell_chemophys,Bassler09}).
Identifying the underlying mechanisms (signaling modality, signaling agent,
signal transport, and so on) as well as the molecules and structures
implementing the mechanisms is no small undertaking. Consequently, experimental
biologists use a combination of prior knowledge and what can only be called
instinct to choose those systems on which to expend effort. Guidance may be
sought from evolutionary developmental biology -- a field that compares the
developmental processes of different organisms to determine their ancestral
relationship and to discover how developmental processes evolved. Insights may
be gained by using statistical machine learning techniques to analyze
heterogeneous data such as the biomedical literature and the output of so-called
``omics'' technologies -- genomics (genes, regulatory, and non-coding
sequences), transcriptomics (RNA and gene expression), proteomics (protein
expression), metabolomics (metabolites and metabolic networks), pharmacogenomics
(how genetics affects hosts' responses to drugs), and physiomics (physiological
dynamics and functions of whole organisms).

Typically, the application of communication theory to biology starts by
selecting a candidate system whose components and operations have been already
elucidated to varying degrees using methods in the experimental and/or
computational biology toolbox \cite{HH4,Long09} and then applying communication
theoretic methods \cite{todd_ITspecial10,wojtek,donjohnson10,emery04,Bassler09}.
However, we believe that communication theory in general and information theory
in particular are not merely system analysis tools for biology.  That is, given
energy constraints and some general physics of the problem, an
information-theoretic treatment can be used to provide outer bounds on
information transfer in a {\em mechanism-blind} manner. Thus, rather than simply
elucidating and quantifying known biology, communication theory can winnow the
plethora of possibilities (or even suggest new ones) amenable to experimental
and computational pursuit. Likewise, general application of
communication-theoretic principles to biology affords a new set of application
areas for communication theorists. Some aspects of the potential for
communication theory as a new lens on biological systems are explored in
\cite{MianRose11}.

In this light, here we devise an abstraction that encompasses a myriad of
biological processes and phenomena, utilize it to devise a simpler model
suitable for communication-theoretic investigations, and analyze the resultant
model using ideas discussed in seemingly unrelated work, namely the capacity of timing
channels \cite{bits-Qs}. Numerous scenarios in biology that involve the
transmission of information can be synthesized and summarized as inscribed
matter is sent by an emitter, moves through a medium, and 
arrives eventually at its destination receptor where it is interpreted.

Scenarios illustrating the complexity and diversity that our abstraction
attempts to capture include the following:
\begin{itemize}
\item
messenger RNA molecules (mRNAs) that are transcribed from the genome
migrate from the nucleus to the cytoplasm where they are translated
by the ribosome into proteins.
\item
Molecules of the neurotransmitter acetylcholine (Ach) that are released
by the presynaptic neuron terminal diffuse through the synaptic cleft
and bind to nicotinic Ach receptors on the motor end plate.
\item
Ions, molecules, organelles, bacteria and viruses that are present in
one cell are shipped through a thin membrane channel (tunneling nanotube)
to the connected cell where they elicit a physiological response.
\item
Membrane-bound vesicles that contain a variety of materials and substances 
translocate through the cytoplasm to the cell membrane where release
their contents into the extracellular environment.
\item
Malignant cells that have escaped the confines of a tissue circulate
through the bloodstream to other sites where they re-penetrate the
vessel walls and can seed a new tumor.
\item
Chemicals factors that are secreted or excreted by an individual travel
outside the body where they are sensed by a member of the same species
triggering a social or behavioral response.
\end{itemize}
Although the abstraction accommodates a wide range of spatiotemporal
scale and types of emitters, inscribed matter, and receptors, it neglects
many biologically important features. For example, the suite of signaling
quanta -- molecules, macromolecular complexes, organelles, cells, and
so on -- that are released is not necessarily the same as that which
reaches the target because some may be changed (eukaryotic mRNAs are
modified post-transcriptionally), some may be removed (Ach
can be degraded by the enzyme Ach-esterase), some never arrives
(the random path produced by diffusion may result in a trajectory that
leads away from the target \cite{eckford1}), and so on. The movement of inscribed matter
may be passive or active, may or may not require energy and so on.

Despite its limitations, the abstraction does embody a number of salient
features. Typically, information is thought to be conveyed via numbers of
signaling quanta (concentration). Thus, what amount to dose-response curves are
the norm for a variety of experimental biology studies \cite{Bassler09} and
clever theoretical workups (e.g., \cite{fekriisit11}).  However, as was shown in
an entirely different domain and unrelated work \cite{bits-Qs}, timing of
emissions could {\em in principle} also convey information. Clearly, this
possibility cannot be ignored if our aim is to attempt to provide bounds on what
``a cell can tell the world.'' Under certain conditions, perhaps timing is a
useful complement to concentration or even essential.  Alternatively, timing
might sometimes be energetically unfavorable and its use unlikely. In either
case, information-theoretic bounds would help guide biological inquiry.

Our emitter-receptor system is also, at least in part, motivated by fundamental
``systems'' problems in biology such as development, wherein undifferentiated
cells are ``told'' what to become by a combination of internal programming and
extracellular milieu signals -- and in turn tell other cells what to become
\cite{nuss}. Thus, communication within and between cells plays a vital role in
the development (embryogenesis), maintenance (tissue homeostasis), subversion
(disorders such as cancer, inflammation, infections) and decline (aging) of
multicellular forms and systems.

Unfortunately, the detailed physics of even this seemingly simple abstraction
are fraught with a variety of complications. As indicated above, free-space
diffusive first passage times are generally not at all well-behaved.  There may
be deletions (a quantum is captured and destroyed by ``lysing'' agents) or the
first passage density may be heavy-tailed to the point that sometimes some of
the inscribed matter may {\em never} arrive at the receptor site
\cite{eckford1}.  Here we will ignore both complications.  Random deletions can
only reduce information transfer, so assuming quanta survive transit provides an
upper bound.  Likewise for heavy-tailed first passage densities, cells emitting
signaling quanta into constrained extracellular (or even more tightly
constrained intracellular) media, arrival with finite mean first passage time
seems reasonable.

However, the most technically difficult complication -- and one which cannot be
ignored -- is quanta indistinguishability.  Which emission corresponds to which
arrival can be ambiguous.  That is, if emissions occur at times $\{T_i\}$ and
the corresponding arrivals occur at $\{ S_i \}$, then all the receiver has
available is $\{ {\Svec}_i \}$, the time-ordered version of the arrivals. Thus,
our major task is to derive expressions for $I(\Svv;\Tmat)$ and thence
$\max_{f_{\Tmat}}I(\Svv;\Tmat)$.

In what follows we first formally define the problem, provide some simplifying
symmetry assumptions, explore their implications and then derive expressions for
the mutual information between quantum launch times and time-ordered quantum
arrival times.  We consider the analytically tractable special case of
exponential first passage, fold in the cost of quantum manufacture and consider
capacity per unit energy (capacity per quantum).  We defer exploration of
physiologically-derived parameters applied to our results for future work.

\section{Problem Definition}
We assume that $M$ identical quanta are emitted at times $\{T_m\}$, $m =
1,2,...,M$.  The duration of quantum $m$'s first-passage between source and
destination is $D_m$. These $D_m$ are assumed i.i.d. with $f_{D_m}(d) = g(d) =
G^{\prime}(d)$ where $g()$ is some causal probability density with mean
$\frac{1}{\lambda}$ and CDF $G()$.  We also assume that $g()$ contains no
singularities.  Thus, the first portion of the channel is modeled as a sum of
random $M$-vectors
\be
\Smat = \Tmat + \Dmat
\ee
for which we have
\be
\label{eq:f_s_def}
f_{\Smat}(\sv)
=
\int_{\bzero}^{\binfty}  f_{\Tmat}(\tv) f_{\Smat|\Tmat}(\sv|\tv)  d \tv
=
\int_{\bzero}^{\sv} f_{\Tmat}(\tv)  \prod_{m=1}^M g(s_m - t_m)  d \tv
=
\int_{\bzero}^{\sv} f_{\Tmat}(\tv)  \gv(\sv - \tv)  d \tv
\ee
where
$$
\gv(\sv - \tv)
=
\prod_{m=1}^M g(s_m - t_m) 
$$
and we impose an emission deadline, $T_m \le \tau$, $\forall m \in \{ 1,2,...,M
\}$.  The associated emission time ensemble probability density $f_{\Tmat}(\tv)$
is assumed causal, but otherwise arbitrary. We define the launch and capture of $M$
quanta is defined as a ``channel use.''  If we assume multiple independent
channel uses, then the usual coding theorems apply \cite{cover} and the
channel's figure of merit is the mutual information between $\Tmat$ and $\Svv$,
$I(\Svv;\Tmat)$.  We will seek to understand the behavior of $I(\Svv;\Tmat)$ and
provide bounds on its maximum an minimum.

\begin{wrapfigure}{r}{2.2in}
\vspace{-0.4in}
\begin{center}
\includegraphics[height=2.0in,width=2.0in]{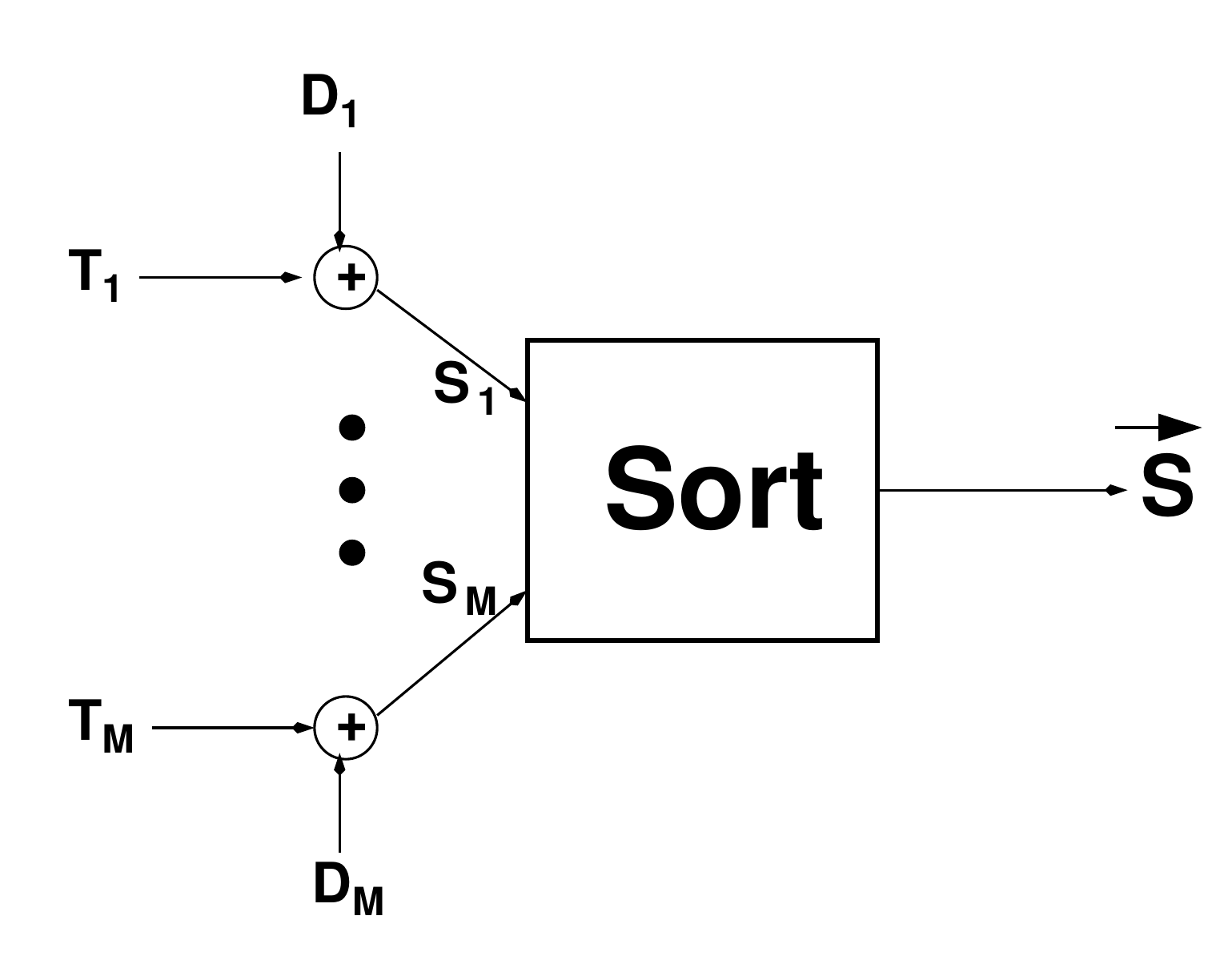}
\end{center}
\vspace{-0.5in}
\caption{\small Quantum release channel with reordering.}
\label{figure:quantumchannel}
\end{wrapfigure}
Had we imposed a mean constraint instead of a deadline, the channel between
$\Tmat$ and $\Smat$ would be parallel version of the model introduced in {\em
  Bits Through Queues} \cite{bits-Qs}.  Even so, since the quanta are identical
we cannot necessarily determine which arrival corresponds to which emission
time.  Thus, the final output of the channel is a reordering of the $\{ s_m \}$
to obtain a set $\{ \vec{s}_m \}$ where $\vec{s}_m \le \vec{s}_{m+1}$,
$m=1,2,..., M-1$. (See FIGURE~\ref{figure:quantumchannel}.) We write this
relationship as
\be
\Svv = P_{\na}(\Smat)
\ee
where $P_k()$, $k=1,2,\cdots,M!$, is a permutation operator and $\na$ is that permutation index which
produces an ordered $\Svv$ from the argument $\Smat$.  Incidentally, we define $P_1()$
as the identity permutation operator, $P_1(\sv) = \sv$.

We note that the event $S_i = S_j$ ($i \ne j$) is of zero measure owing to the
no-singularity assumption on $g()$, Thus, for analytic convenience we will assume
that $f_{\Smat}(\sv) = 0$ whenever two or more of the $s_m$ are equal and
therefore that the $\{ \vec{s}_m \}$ are strictly ordered wherever $f_{\Svv}() \ne 0$
(i.e., $\vec{s}_m < \vec{s}_{m+1}$).

Thus, the density $f_{\Svv}(\svv)$ can be found by ``folding'' the density
$f_{\Smat}(\sv)$ about the hyperplanes described by one or more of the $s_m$
equal until the resulting probability density is nonzero only on the region
where $s_m < s_{m+1}$, $m=1,2,...,M-1$.  Analytically we have
\be
\label{eq:svv_def}
f_{\Svv}(\sv)
=
\twodef{\sum_{n=1}^{M!}
f_{\Smat}(P_{n}(\sv))}{s_1 < s_2 < \cdots < s_m}
{0}{\mbox{otherwise}}
\ee

We can likewise describe $f_{\Svv|\Tmat}(\sv|\tv)$ as
\be
\label{eq:svvT_def0}
f_{\Svv|\Tmat}(\sv|\tv)
=
\twodef{\sum_{n=1}^{M!}
f_{\Smat|\Tmat}(P_{n}(\sv)|\tv)}{s_1 < s_2 < \cdots < s_m}
{0}{\mbox{otherwise}}
\ee
which to emphasize the assumed causality of $g()$ we rewrite as
\be
\label{eq:svvT_def}
f_{\Svv|\Tmat}(\sv|\tv)
=
\twodef{\sum_{n=1}^{M!}
\gv(P_n(\sv) - \tv) \uv(P_n(\sv) - \tv)
}{s_1 < s_2 < \cdots < s_m}
{0}{\mbox{otherwise}}
\ee
where
$$
\uv(P_n(\sv) - \tv)
=
\prod_{m=1}^M
u([P_n(\sv)]_m - t_m)
$$
and $u()$ is the usual unit step function.

When $g(d) = \lambda e^{-\lambda d}u(d)$, the conditional distribution
on the ordered output $\Svv$ takes the particularly simple form
\be
\label{eq:econd}
f_{\Svv|\Tmat}(\sv|\tv)
=
\lambda^M
e^{ - \lambda \displaystyle{\sum_{i=1}^M} (s_i - t_i)}
\left (
\sum_{n=1}^{M!}
\uv(P_n(\sv) - \tv)
\right )
\ee
for $s_1 < s_2 < \cdots < s_m$.  It is worth mentioning explicitly that
\equat{econd} {\em does not assume} $s_i \ge t_i$ as might be implicit in \equat{f_s_def}.

With these preliminaries done, we can now begin to examine the mutual
information between $\Tmat$, $\Smat$ and $\Svv$.  

\section{Mutual Information Between $\Tmat$ and $\Svv$}
\label{sect:mutualinfo}
The mutual information between $\Tmat$ and $\Smat$ is
\be
\label{eq:unorderedMI}
I(\Smat; \Tmat)
=
h(\Smat) - h(\Smat|\Tmat)
=
M \left ( h(S) - h(S|T) \right )
\ee
Since the $S_i$ given the $T_i$ are mutually independent, $h(\Smat|\Tmat)$ does not
depend on $f_{\Tmat}(\tv)$.  Thus, maximization of \equat{unorderedMI} is simply
a maximization of the marginal $h(S)$ over the marginal $f_T(t)$, a problem
explicitly considered and solved for a mean $T_m$ constraint in \cite{bits-Qs}.

The corresponding expression for the mutual information between $\Tmat$ and $\Svv$ is
\be
\label{eq:orderedMI}
I(\Svv; \Tmat)
=
h(\Svv) - h(\Svv|\Tmat)
\ee
Unfortunately, $h(\Svv|\Tmat)$ now {\em does} depend on the input distribution
and the optimal form of $h(\Svv)$ is non-obvious.  So, rather than attempting a
brute force optimization of \equat{orderedMI} by deriving order distributions
\cite{eckford1}, we first invoke simplifying symmetries.

Consider that an emission vector $\tv$ and any of its permutations $P_n(\tv)$
produce statistically identical outputs $\Svv$ owing to the reordering operation
as depicted FIGURE~\ref{figure:quantumchannel}.  Thus, any $f_{\Tmat}()$ which
optimizes \equat{orderedMI} can be ``balanced'' to form an optimizing input
distribution which obeys
\be
\label{eq:Tsymm}
f_{\Tmat} (\tv)
=
f_{\Tmat} (P_n(\tv))
\ee
for $n=1,2,...,M!$ and $P_n()$ the previously defined permutation
operator. We will therefore restrict our search to ``hyper-symmetric'' densities
$f_{\Tmat}(\tv)$ as defined by \equat{Tsymm}.

If we assume $f_{\Tmat}()$ is hyper-symmetric, then it is easy to show that
$f_{\Smat}()$ must also be hyper-symmetric.  From \equat{f_s_def} we have
$$
f_{\Smat}(P_n(\sv))
=
\int_{\bzero}^{P_n(\sv)} f_{\Tmat}(\tv)  \gv(P_n(\sv) - \tv)  d \tv
$$
If we define $\tv^{\prime} = P_n^{-1}(\tv)$ then we can write
$$
f_{\Smat}(P_n(\sv))
=
\int_{\bzero}^{\sv} f_{\Tmat}(P_n^{-1}(\tv^{\prime}))  \gv(\sv - \tv^{\prime})  d \tv^\prime
=
\int_{\bzero}^{\sv} f_{\Tmat}(\tv^{\prime})  \gv(\sv - \tv^{\prime})  d \tv^\prime
=
f_{\Smat}(\sv)
$$

The hyper-symmetry of $f_{\Smat}(\sv)$ leads to a simple expression for
$f_{\Svv}(\sv)$.  First we define ${\cal S}_1$ as the region in $\sv$-space for
which $s_1 < s_2 < \cdots < s_m$.  Similarly define disjoint regions ${\cal
  S}_n$ as those for which if $\sv \in {\cal S}_n$ then $P_n(\sv) \in {\cal
  S}_1$.  That is, ${\cal S}_n$ is the region in $\sv$-space in which
application of permutation operator $P_n()$ orders the components from smallest
to largest.

Following \equat{svv_def} we have
$$
f_{\Svv}(\sv)
=
M! f_{\Smat}(\sv)
$$
for $\sv \in {\cal S}_1$.  We can then write
$$
h(\Svv)
=
-\int_{{\cal S}_1}
M! f_{\Smat}(\sv) \log \left (M! f_{\Smat}(\sv)  \right )
d\sv
=
-M! 
\int_{{\cal S}_1}
f_{\Smat}(\sv) \log f_{\Smat}(\sv)
d\sv
-
\log M!
$$
But since $f_{\Smat}(\sv)$ is hyper-symmetric, we also have
$$
h(\Svv)
=
-\sum_{n=1}^{M!}
\int_{{\cal S}_n}
f_{\Smat}(P_n(\sv)) \log f_{\Smat}(P_n(\sv)) d \sv
- \log M!
$$
which
becomes
\be
\label{eq:hyperS}
h(\Svv)
=
-
\int_{\bzero}^{\binfty}
f_{\Smat}(\sv) \log f_{\Smat}(\sv) d \sv
- \log M!
=
h(\Smat) - \log M!
\ee
We state this result as a theorem.
\begin{theorem}
\thmlabel{orderedS}
If $f_{\Tmat}()$ is a hyper-symmetric probability density function on emission
times $\{T_m\}$, $m=1,2,..,M$, and the first passage density is non-singular,
then the entropy of the size-ordered outputs $\Svv$ is
$$
h(\Svv) = h(\Smat) - \log M!
$$
\end{theorem}

Next we turn to $h(\Svv|\Tmat)$. A zero-measure edge-folding argument on the
conditional density is not easily applicable here, so we resort to some sleight
of hand.  As before we define $\Omega$ as the permutation index number that
produces an ordered output from $\Smat$.  That is, $P_{\Omega}(\Smat) = \Svv \in
{\cal S}_1$.  Specification of the random tuple $(\Omega, \Svv)$ is equivalent
to specifying $\Smat$ and {\em vice versa}.  Just as in our derivation of
$h(\Svv)$, this equivalence requires that we exclude the zero-measure ``edges''
and ``corners'' of the density where two or more of the $\svv_i$ are equal.

We then have,
\be
\label{eq:jointdiscretedef}
h(\Smat|\Tmat)
=
h(\Omega, \Svv|\Tmat)
=
h(\Svv|\Tmat)
+
H(\Omega|\Svv, \Tmat)
\ee
which also serves as a definition for the entropy of a joint mixed distribution
($\Omega$ is discrete while $\Sv$ is continuous).  We then rearrange
\equat{jointdiscretedef} as
\be
\label{eq:Homega}
h(\Svv|\Tmat)
=
h(\Smat|\Tmat)
-
H(\Omega|\Svv, \Tmat)
\ee
$H(\Omega|\Svv, \Tmat)$ is the uncertainty about which $S_m$ corresponds to which $\vec{S}_m$ given both $\Tmat$ and $\Svv$, and we note that 
\be
\label{eq:HOeasybounds}
0 \le H(\Omega|\Svv, \Tmat) \le \log M!
\ee
with equality on the right for any density where all the $T_m$ are equal.

We can then, after assuming that $f_{\Tmat}()$ is hyper-symmetric, write the
ordered mutual information in an intuitively pleasing form:
\begin{theorem}
\thmlabel{Isvv}
\be
\label{eq:orderedMI_decomp}
I(\Svv; \Tmat)
=
I(\Smat;\Tmat)
-
\left ( \log M! - H(\Omega|\Svv,\Tmat) \right )
\ee
\end{theorem}
That is, an information degradation of size $\log M! - H(\Omega|\Svv,\Tmat) \le
0$ is introduced by the sorting operation.

Since $h(\Smat|\Tmat)$ is a constant with respect to $f_{\Tmat}(\tv)$, 
maximization of mutual the information in \equat{orderedMI_decomp} requires we maximize
the expression
\be
\label{eq:hH}
\begin{array}{rcl}
h(\Smat) + H(\Omega|\Svv, \Tmat)
 & = & -
\displaystyle{\int_{\bzero}^\binfty}
\displaystyle{\int_{\tv}^\binfty
f_{\Tmat}(\tv)
\gv(\sv-\tv) \log
\overbrace{
\left (
\int_{\bzero}^{\sv}
f_{\Tmat}(\vv) g(\sv - \vv) d\vv
\right )}^{f_{\Smat}(\sv)}
d \sv  d \tv}\\
 & + & \displaystyle{\int_{\bzero}^\binfty
f_{\Tmat}(\tv)
H(\Omega|\Svv, \tv)
d \tv
}
\end{array}
\ee
with respect to $f_{\Tmat}(\tv)$.

Mutual information is convex in $f_{\Tmat}(\tv)$ and the space ${\cal
  F}_{\Tmat}$ of feasible hyper-symmetric $f_{\Tmat}(\tv)$ is convex. That is,
for any two hyper-symmetric probability functions $f_{\Tmat}^{(1)}$ and
$f_{\Tmat}^{(2)}$ we have
\be
\label{eq:convexf_T}
\kappa f_{\Tmat}^{(1)}(\tv) + (1-\kappa) f_{\Tmat}^{(2)}(\tv)
\in {\cal F}_{\Tmat}
\ee
where $0 \le \kappa \le 1$.  Thus, we can in principle apply variational
\cite{hild} techniques to find that hyper-symmetric $f_{\Tmat}()$ which attains
the unique maximum of \equat{orderedMI}. However, in practice, direct
application of this method can lead to grossly infeasible $f_{\Tmat}()$,
implying that the optimizing $f_{\Tmat}()$ lies along some ``edge'' or in some
``corner'' of the convex search space.

To proceed, we must first understand the component parts of the optimization, in
particular $H(\Omega|\Svv,\Tmat)$ and its relationship to $h(\Svv)$.  But first
the following property of expectations of hyper-symmetric functions over
hyper-symmetric random variables will later prove useful.  Suppose $Q(\xv)$ is a
hyper-symmetric function and $\Xmat$ is a hyper-symmetric random vector.  Then,
when $\vec{\Xmat}$ is the ordered version of random vector $\Xmat$ we have \be
\label{eq:hyperexpect}
E_{\Xvv} \left [Q(\Xvv) \right ]
=
E_{\Xmat} \left [Q(\Xmat) \right ]
\ee

\subsection{$H(\Omega|\Svv,\tv)$}
The optimization stated in \equat{hH} hinges on specification of
$H(\Omega|\Svv,\Tmat)$.  We first consider $H(\Omega|\svv,\tv)$, the
admissible-permutation entropy given specific $\svv$ and $\tv$.  Given $\tv$, the
probability that $\Smat$ produced $\Svv$ is
\be
\label{eq:Pst}
\mbox{Prob}(\Omega| \svv, \tv)
=
\frac{f_{\Smat|\Tmat}(\svv| \tv)}
{\displaystyle{\sum_{n=1}^{M!}}
f_{\Smat|\Tmat}(P_n(\svv)| \tv)}
\ee
where $\svv = P_{\na}(\sv)$.  Owing to the causality of $g()$, some
permutations will have zero probability since the specific $\svv$ and $\tv$ may
render them impossible.

Using \equat{svvT_def}, the definition of entropy and \equat{Pst} we have
\be
\label{eq:HstG}
H(\Omega|\svv,\tv)
=
-
\sum_{n=1}^{M!}
\left  [
\frac{\gv(P_n(\svv) - \tv)}{\displaystyle{\sum_{j=1}^{M!}}\gv(P_j(\svv) - \tv)}
\right ]
\log
\left [
\frac{\gv(P_n(\svv) - \tv)}{\displaystyle{\sum_{j=1}^{M!}}\gv(P_j(\svv) - \tv)}
\right ]
\ee
and as might be imagined, \equat{HstG} does not in general produce a closed form.

However, for exponential $g()$ we can use \equat{econd} to simplify \equat{Pst}
as
\be
\label{eq:PstE}
\mbox{Prob}(\Omega = k| \svv, \tv)
=
\frac{\uv(\svv- \tv)}
{\displaystyle{\sum_{n=1}^{M!}}
\uv(P_n(\svv)- \tv)}
\ee
which is a uniform probability mass function with $\sum_{n=1}^{M!}\uv(P_n(\svv)-
\tv)$ elements. Thus, we can write
\be
H_e(\Omega|\svv,\tv)
=
\log 
\displaystyle{\sum_{n=1}^{M!}}
\uv(P_n(\svv)- \tv)
\ee
The summation is the number of admissible permutations given $\svv$ and $\tv$,
and constitutes an upper bound for all possible causal first-passage time
densities, $g()$. In addition, the exponential first passage time density is the
{\em only} density which maximizes $H_e(\Omega|\svv,\tv)$.
We state the result as a theorem:
\begin{theorem}
\thmlabel{Homegaupperbound}
If we define
$$
\left |
\Omega
\right |_{\svv,\tv}
=
\sum_{n=1}^{M!}
\uv(P_n(\svv)- \tv)
$$
then
$$
H(\Omega|\svv,\tv)
\le
\log 
\left |
\Omega
\right |_{\svv,\tv}
$$
with equality {\bf iff} $g()$ is exponential.
\end{theorem}
\begin{Proof}(\Thmref{Homegaupperbound}) Although \equat{PstE} constitutes a proof that the
  exponential first passage time density maximizes $H_e(\Omega|\svv,\tv)$, we can also prove the
  {\bf iff} result directly. Consider that the probability mass (PMF) function of \equat{Pst} can be written as
$$
\mbox{Prob}(\Omega = k| \svv, \tv)
=
\frac{\gv(P_k(\svv) - \tv)}{\displaystyle{\sum_{j=1}^{M!}}\gv(P_j(\svv) - \tv)}
$$
This PMF is uniform {\bf iff}
\be
\label{eq:permute}
\gv(P_n(\svv) - \tv)
=
\gv(P_k(\svv) - \tv)
\ee
for all $n$ and $k$ where $P_n(\svv)$ and $P_k(\svv)$ are causal with respect to
$\tv$.  That is, the pairs $(P_n(\svv), \tv)$ and $(P_k(\svv), \tv)$ are {\em
  admissible}. Since the maximum number of non-zero probability $\Omega$ is
exactly the cardinality of admissible $(P_n(\svv), \tv)$, any density which
produces a uniform PMF over $\Omega$ thereby maximizes $H(\Omega|\svv,\tv)$.

We then note that any given permutation of a list can be achieved by sequential
pairwise swapping of elements.  Thus, \equat{permute} is satisfied {\bf iff}
\be
\label{eq:product}
g(x_1 - t_1)g(x_2-t_2)
=
g(x_2 - t_1)g(x_1-t_2)
\ee
$\forall$ admissible $\{ (x_1,x_2)$,  $(t_1,t_2)  \}$.  Rearranging \equat{product} we have
$$
\frac{g(x_1-t_1)}{g(x_1-t_2)}
=
\frac{g(x_2-t_1)}{g(x_2-t_2)}
$$
which implies that
$$
\frac{g(x-t_1)}{g(x-t_2)}
=
\mbox{Constant w.r.t. $x$}
$$

Differentiation with respect to $x$ yields
$$
\frac{g^\prime(x-t_1)}{g(x-t_2)}
-
\frac{g(x-t_1)g^\prime(x-t_2)}{g^2(x-t_2)}
=0
$$
which we rearrange to obtain
$$
\frac{g^\prime(x-t_1)}{g(x-t_1)}
=
\frac{g^\prime(x-t_2)}{g(x-t_2)}
$$
which further implies that
$$
\frac{g^\prime(x-t_1)}{g(x-t_1)}
=
c
$$
whose only solution is
$$
g(x) \propto e^{cx}
$$
Thus, exponential $g()$ is the only first passage time density that can produce a
maximum cardinality uniform distribution over $\Omega$ given $\svv$ and $\tv$ --
which completes the proof.
\end{Proof}

Now consider that $\left | \Omega \right |_{\svv,\tv}$ is hyper-symmetric --
invariant under any permutation of its arguments $\svv$ or $\tv$. That is,
$$
\sum_{n=1}^{M!}
\uv(P_n(\svv) - \tv)
=
\sum_{n=1}^{M!}
\uv(P_n(\svv)- \tvv)
=
\sum_{n=1}^{M!}
\uv(P_{n}(\sv)- \tvv)
=
\sum_{n=1}^{M!}
\uv(P_n(\sv)- \tv)
$$
because the summation is over all $M!$ permutations.  Therefore, 
\be
\label{eq:symmetry}
\left |
\Omega
\right |_{\svv,\tv}
=
\left |
\Omega
\right |_{\svv,\tvv}
=
\left |
\Omega
\right |_{\sv,\tvv}
=
\left |
\Omega
\right |_{\sv,\tv}
\ee

We now enumerate admissible permutations.
 Owing to \equat{symmetry} we can assume ordered $\tv$ with no loss of
generality. So, let us define ``bins'' ${\cal B}_k = \{t|
t \in [t_k, t_{k+1}) \}$, $k=1,2,...,M$ ($t_{M+1} \equiv \infty$) and let
$b_m = 1,2,..., M$ be the bin in which $\svec_m$ appears ($\svec_m \in {\cal
  B}_{b_m}$).  We then define $\sigma_m$ as bin occupancies such that $\sigma_m =
q$ if there are exactly $q$ arrivals $s_i \in {\cal B}_m$.  The benefit of this
approach is that the $\sigma_m$, do not depend on whether $\svv$ or $\sv$ is used.
Thus, expectations can be taken over $\Smat$ whose components are mutually
independent given the $\tv$.

To calculate $\left | \Omega
\right |_{\svv,\tv}$ we start by defining
$$
\eta_m = \sum_{j=1}^{m} \sigma_j
$$
Clearly $\eta_m$ is monotonically increasing in $m$ with $\eta_0 = 0$ and
$\eta_M = M$.  We then observe that the $\sigma_m$ arrivals on $[t_m,t_{m+1})$ can be
assigned to any of the $t_1, t_2, ..., t_m$ known emission times {\em except}
for those $\eta_{m-1}$ previously assigned.  The number of possible new assignments
is $(m-\eta_{m-1})!/(m - \eta_m)!$ which leads to
\be
\label{eq:Omegacardb}
\left |
\Omega
\right |_{\sv,\tv}
=
\prod_{m=1}^M
\frac{(m - \eta_{m-1})!}{(m - \eta_{m})!}
=
\prod_{m=1}^{M-1}
(m + 1 - \eta_m)
\ee

We then define the random variable
$$
X_i^{(m)}
=
\twodef{1}{S_i < t_{m+1}}{0}{\mbox{otherwise}}
$$
for $i=1,2,...m$.  The PMF of $X_i^{(m)}$ is then
$$
p_{X_i^{(m)}}(x)
=
\twodef{G(t_{m+1} - t_i)}{x = 1}{\bar{G}(t_{m+1} - t_i)}{x = 0}
$$
where as previously defined, $G()$ is the CDF of the causal first passage density $g()$ and $\bar{G}() = 1 - G()$ is its CCDF.  We can then write
$$
\eta_m = \sum_{i=1}^m  X_i^{(m)}
$$
and thence via \equat{Omegacardb},
\be
\label{eq:HSMAT}
E_{\Smat|\tv} [ \left |
\Omega
\right |_{\Smat,\tv}]
=
E_{\Smat|\tv}  \left [\sum_{m=1}^M \log (m + 1 - \eta_m) 
\right ]
\ee
Since an expectation of a sum is the sum of the expectations, let us consider
\be
E_{\Smat|\tv} [\log (m + 1 - \eta_m) ]
=
\sum_{\xv}
\log (m + 1 - \sum_{i=1}^m x_i)
\prod_{j=1}^m
G^{x_j}(t_{m+1} - t_j)
\bar{G}^{1 -x_j}(t_{m+1} - t_j)
\ee
where $\xv$ is implicitly an $m-$ary binary vector.

We now find it convenient to define
$\bar{X}_i = 1 - X_i$ which allows us to define $\bar{\eta}_m = m - \eta_m$
and thence
\be
\label{eq:Omegacardbbar}
\left |
\Omega
\right |_{\sv,\tv}
=
\prod_{m=1}^{M-1}
(1 + \bar{\eta}_m)
\ee
and 
\be
\label{eq:Homega_etabar}
E_{\Smat|\tv} [\log (1 + \bar{\eta}_m) ]
=
\sum_{\bar{\xv}}
\log (1 + \sum_{i=1}^m \bar{x}_i)
\prod_{j=1}^m
\bar{G}^{\bar{x}_j}(t_{m+1} - t_j)
G^{1 -\bar{x}_j}(t_{m+1} - t_j)
\ee

We can now define
\be
\label{eq:Hexpurg}
\Hup(\tv)
=
\sum_{m=1}^{M-1}
\sum_{\bar{\xv}}
\log (1 + \sum_{i=1}^m \bar{x}_i)
\prod_{j=1}^m
\bar{G}^{\bar{x}_j}(t_{m+1} - t_j)
G^{1 -\bar{x}_j}(t_{m+1} - t_j)
\ee
which can be rearranged as
\be
\label{eq:Hexpurg3}
\Hup(\tv)
=
\sum_{\ell=1}^{M-1}
\log(1 + \ell)
\sum_{m=\ell}^{M-1}
\sum_{|\bar{\xv}| = \ell}
\prod_{j=1}^m
\bar{G}^{\bar{x}_j}(t_{m+1} - t_j)
G^{1 -\bar{x}_j}(t_{m+1} - t_j)
\ee
We will also later find it useful to define
\be
\label{eq:thetadef}
\Theta_{m,\ell}(\tv)
\equiv
\sum_{|\bar{\xv}| = \ell}
\prod_{j=1}^m
\bar{G}^{\bar{x}_j}(t_{m+1} - t_j)
G^{1 -\bar{x}_j}(t_{m+1} - t_j)
\ee
which produces
\be
\label{eq:Hexpurg4}
\Hup(\tv)
=
\sum_{\ell=1}^{M-1}
\log(1 + \ell)
\sum_{m=\ell}^{M-1}
\Theta_{m,\ell}(\tv)
\ee
which after defining
$$
\Gamma_{M,\ell}
=
\sum_{m=\ell}^{M-1}
\Theta_{m,\ell}
$$
and
$$
\Delta\Gamma_{M\ell}
=
\Gamma_{M,\ell} -  \Gamma_{M,\ell+1}
$$
can be rewritten as
\be
\label{eq:Hexpurg5}
\Hup(\Tmat)
=
\sum_{\ell=1}^{M-1}
\Delta\Gamma_{M\ell}
\log (\ell + 1)!
\ee
where, once again, we have assumed that $t_1 \le t_2 \le \cdots \le t_m$.

Finally, via \equat{symmetry}, \equat{HSMAT} and the definition of \equat{Hexpurg} in conjunction with \Thmref{Homegaupperbound} we have
\begin{theorem}
\thmlabel{Hbound}
If we define
$$
\Hup(\Tmat)
\equiv
E_{\Tvv} \left [ \Hup(\Tvv) \right ]
$$
then since
$$
H(\Omega|\Svv,\tv) 
\le
\Hup(\tv)
$$
we have
$$
H(\Omega|\Svv,\Tmat) 
\le
\Hup(\Tmat)
$$
with equality {\bf iff} the first-passage time density
$g()$ is exponential.
\end{theorem}

\subsection{$\Hup(\Tmat)$}
In principle, we could derive $\Hup(\Tmat)$ by taking the expectation of
\equat{Hexpurg4} with respect to ordered emission times.  Although we can do just
that for numerical calculations, direct {\em analytic} evaluation of
$\Hup(\Tmat)$ requires we derive joint order densities for the $\Tmat$, a
difficult task in general.  Thus, for analytic simplicity we will take advantage
of emission time distribution hypersymmetry and derive only univariate order
densities.

That is, the sum over all permutations of binary vector $\bar{\xv}$ in the
definition of $\Theta_{m,\ell}(\tv)$ renders it hypersymmetric in $t_1 , ... ,
t_m$ given the $(m+1)^{\mbox{st}}$ smallest emission time $\tvec_{m+1}$ which for
clarity we denote with the over-arrow notation.  Therefore, by \equat{hyperexpect}
we have
\be
\label{eq:Hhyper}
E_{\Tmat} \left   [
\Theta_{m,\ell}(\Tmat)
\right ]
=
E_{\Tvec_{m+1}} \left   [
E_{T_1,...,T_m|\Tvec_{m+1}}
\left [
\Theta_{m,\ell}(T_1,...T_m,\Tvec_{m+1})
\right ]
\right ]
\ee

The CDF of the $(m+1)^{\mbox{st}}$ smallest emission time is
\be
\label{eq:ftvm}
F_{\vec{T}_{m+1}} (t_{m+1})
=
1 
-
\sum_{k=0}^{m}
{M \choose k}
\underbrace{
\int_0^{t_{m+1}}
\cdots
\int_0^{t_{m+1}}
}_{\mbox{$k$}}
\underbrace{
\int_{t_{m+1}}^{\infty}
\cdots
\int_{t_{m+1}}^{\infty}
}_{\mbox{$M-k$}}
f_{\Tmat}(\tv)
dt_M
\cdots
dt_{k+1}
dt_{k}
\cdots
dt_{1}
\ee
and likewise, the CDF of the smallest {\em unordered} $T_1,...,T_m$ given $\vec{T}_{m+1}$ is
$$
F_{T_1,...,T_m|\vec{T}_{m+1}} (t_1,...,t_m|t_{m+1})
=
\frac{F_{T_1,...,T_m} (t_1,...,t_m)}{F_{T_1,...,T_m} (t_{m+1},...,t_{m+1})}
$$
$\forall t_j \le t_{m+1}$ where $j=1,...,m$.

Therefore, by the hypersymmetry of $\bar{\Theta}_{m,\ell}$ in $t_1,...,t_m$ we may write
$$
\bar{\Theta}_{m,\ell}
=
\int_0^{\infty}
f_{\vec{T}_{m+1}}(t_{m+1})
{\bf \int}_{\bzero}^{{\bf t_{m+1}}}
\frac{f_{T_1,...,T_m}(t_1,...,t_m)}{F_{T_1,...,T_m}(t_{m+1},...,t_{m+1})}
B(m,\ell,\tv)
dt_1
...
dt_{m+1}
$$
where
$$
B(m,\ell,\tv)
\equiv
{m \choose \ell}
\prod_{j=1}^{\ell}
\bar{G}(t_{m+1} - t_j)
\prod_{k=\ell+1}^m
G(t_{m+1} - t_k)
$$
and thence
\be
\label{eq:He1}
\Hup(\Tmat)
=
\sum_{\ell=1}^{M-1}
\log(1 + \ell)
\sum_{m=\ell}^{M-1}
\bar{\Theta}_{m,\ell}
\ee
And similar to the derivation of \equat{Hexpurg5}, we define
$$
\bar{\Gamma}_{M,\ell}
=
\sum_{m=\ell}^{M-1}
\bar{\Theta}_{m,\ell}
$$
and then
$$
\overline{\Delta\Gamma}_{M\ell}
=
\bar{\Gamma}_{M,\ell} -  \bar{\Gamma}_{M,\ell+1}
$$
to express $\Hup(\Tmat)$ as
\be
\label{eq:He2}
\Hup(\Tmat)
=
\sum_{\ell=1}^{M-1}
\overline{\Delta\Gamma}_{M\ell}
\log (\ell + 1)!
\ee

\section{IID $\Tmat$}
Our attempts at direct optimization of \equat{hH} have not yielded a closed
form.  The key problem is that $h(\Smat)$ and $H(\Omega|\Tmat,\Svv)$ are
``conflicting'' quantities with respect to $f_{\Tmat}()$.  That is, independence
of the $T_m$ favors larger $h(\Smat)$ while tight correlation of the $T_m$ (as
in $T_i=T_j$, $i,j = 1,2,...,M$) produces the maximum $H(\Omega|\Svv,\Tmat) =
\log M!$.  It is this tension which leads to grossly infeasible $f_t()$ (with
high order singularities) when applying standard Lagrange-Euler variational
optimization methods to \equat{hH}.  In short, a closed-form upper bound tighter
than that provided by the data processing theorem \cite{cover}
\be
\max_{f_\Tmat()}I(\Svv;
\Tmat) \le \max_{f_\Tmat()}I(\Smat,\Tmat)
\ee

has so far eluded us.

We therefore derive expressions for $I(\Svv;\Tmat)$ when the $\Tmat$ are IID --
as they must be to maximize $I(\Smat;\Tmat)$.  Such an assumption has some
grounding in the biology of signaling in that quanta (signaling molecules) are
often emitted from physically distinct and separate repositories (vesicles).
Thus, coordinating emission times could add complexity to the release mechanism.
By deriving expressions for $I(\Svv;\Tmat)$ given IID $\Tmat$ -- which form
lower bounds for $\max_{f_{\Tmat}}I(\Svv;\Tmat)$ in the case of exponential
first passage times -- we may provide insight for when the machinery necessary for
tightly coordinated emissions is a worthwhile investment.

\subsection{$\Hup(\Tmat)$ and General IID $\Tmat$}
From the definition of $\theta_{m,\ell}()$ in \equat{thetadef} we obtain
\be
E \left [
\Theta_{m,\ell}(\tv)
\right ]
=
E_{\Tvec_{m+1}} \left [
{m \choose \ell}
E_{T \le \Tvec_{m+1}}^{\ell} \left [
\bar{G}(\Tvec_{m+1} - T)
\right ]
E_{T \le \Tvec_{m+1}}^{m-\ell} \left [
(1 - \bar{G}(\Tvec_{m+1} - T))
\right ]
\right ]
\ee
for IID $\Tmat$.  From the definition of $F_{\Tvec{m+1}}()$ in \equat{ftvm}
we obtain (again for IID $\Tmat$)
\be
f_{\Tvec_{m+1}}(t)
=
\frac{d}{dt}
\left [
1 
-
\sum_{k=0}^{m}
{M \choose k}
F_T^k(t)(1-F_T(t))^{M-k}
\right ]
\ee
which after rearranging as a telescoping sum simplifies to
\be
\begin{array}{rcl}
f_{\Tvec_{m+1}}(t) & = & \sum_{k=0}^{m}
(M-k)
{M \choose k}
f_T(t)F_T^k(t)(1-F_T(t))^{M-k-1}\\
 & - & 
\sum_{k=0}^{m-1}
(k+1)
{M \choose {k+1}}
f_T(t)F_T^{k}(t)(1-F_T(t))^{M-k-1}
\end{array}
\ee
which further simplifies to
\be
f_{\Tvec_{m+1}}(t)
=
(m+1)
{M \choose {m+1}}
f_T(t)F_T^m(t)(1-F_T(t))^{M-m-1}
\ee
If we then define
$$
\phi(t)
=
\int_0^t
f_T(x) \bar{G}(t-x)
dx
$$
we obtain
$$
\int_0^t
f_T(x) (1- \bar{G}(t-x))
dx
=
F_T(t) - \phi(t)
$$
so we can write
$$
\bar{\theta}_{m,\ell}
=
(m+1)
{M \choose {m+1}}
{m \choose {\ell}}
\int_0^{\infty}
f_T(t)(1-F_T(t))^{M-m-1}
\phi^{\ell} (t)
(F_T(t)-\phi(t))^{m-\ell} 
dt
$$
and then as
\be
\label{eq:thetaml}
\bar{\theta}_{m,\ell}
=
M
{{M -1} \choose \ell}
{{M - \ell - 1} \choose {m- \ell}}
\int_0^{\infty}
f_T(t)(1-F_T(t))^{M-m-1}
\phi^{\ell} (t)
(F_T(t)-\phi(t))^{m-\ell} 
dt
\ee

To evaluate \equat{He2} we now compute
$$
\bar{\Gamma}_{M,\ell}
=
\sum_{m=\ell}^{M-1}
\bar{\theta}_{m,\ell}
=
M
{{M -1} \choose \ell}
\int_0^{\infty}
f_T(t)
\left [
\begin{array}{c}
{\displaystyle
\sum_{m=\ell}^{M-1}}
{{M - \ell - 1} \choose {m- \ell}}
\left ( \frac{F_T(t)-\phi(t)}{1 - F_T(t)} \right )^{m}\\
\times
(1- F_T(t))^{M-1}
\left ( \frac{\phi(t)}{F_T(t) - \phi(t)} \right )^{\ell}\\
\end{array}
\right ]
dt
$$
which we rewrite as
$$
M
{{M -1} \choose \ell}
\int_0^{\infty}
f_T(t)
\left [
\begin{array}{c}
{\displaystyle\sum_{m=0}^{M-1-\ell}}
{{M - \ell - 1} \choose {m}}
\left ( \frac{F_T(t)-\phi(t)}{1 - F_T(t)} \right )^{m+\ell}\\
\times (1- F_T(t))^{M-1}
\left ( \frac{\phi(t)}{F_T(t) - \phi(t)} \right )^{\ell}
\end{array}
\right ]
dt
$$
We consolidate the binomial sum to obtain
$$
M
{{M -1} \choose \ell}
\int_0^{\infty}
f_T(t)
\left [
\begin{array}{c}
\left ( \frac{F_T(t)-\phi(t)}{1 - F_T(t)} \right )^{\ell}
\left ( \frac{1-\phi(t)}{1 - F_T(t)} \right )^{M-1-\ell}\\
(1- F_T(t))^{M-1}
\left ( \frac{\phi(t)}{F_T(t) - \phi(t)} \right )^{\ell}
\end{array}
\right ]
dt
$$
which reduces to
\be
\label{eq:gammaML}
\bar{\Gamma}_{M,\ell}
=
M
{{M-1} \choose \ell}
\int_0^{\infty}
f_T(t)
\phi^{\ell}(t)
\left ( 1 -\phi(t) \right )^{M-1-\ell}
dt
\ee
for $\ell = 1,2,...,M-1$.

Now consider the integrand of the difference $\bar{\Gamma}_{M,\ell} -
\bar{\Gamma}_{M,\ell+1}$ where we drop the $t$ dependence for notational convenience
$$
M
{{M-1} \choose \ell}
\phi^{\ell}
\left ( 1 -\phi \right )^{M-\ell-1}
-
M
{{M-1} \choose {\ell+1}}
\phi^{\ell+1}
\left ( 1 -\phi \right )^{M-\ell-2}
$$
We can rewrite this expression as
$$
M
\phi^{\ell}
\left [
{{M-1} \choose \ell}
+
\sum_{r=1}^{M-\ell -1}
(-1)^r
\phi^r
\left [
{{M-1} \choose \ell}
{{M-\ell - 1} \choose r} 
+
{{M-1} \choose {\ell+1}}
{{M-\ell - 2} \choose {r-1}}
\right ]
\right ]
$$
which after consolidating terms becomes
$$
M
\phi^{\ell}
\left [
{{M-1} \choose \ell}
+
\frac{1}{M}
{{M} \choose {\ell+1}}
\sum_{r=1}^{M-\ell -1}
(-1)^r
{{M-\ell -1} \choose {r}}
(\ell+r+1)
\phi^r
\right ]
$$
Extending the sum to $r=0$ and subtracting the $r=0$ term produces
$$
{{M} \choose {\ell+1}}
\sum_{r=0}^{M-\ell -1}
(-1)^r
{{M-\ell -1} \choose {r}}
(\ell+r+1)
\phi^{r+\ell}
$$
so that
\be
\label{eq:Gammadiffgeneral}
\overline{\Delta\Gamma}_{M,\ell} 
=
{{M} \choose {\ell+1}}
\sum_{r=0}^{M-\ell -1}
(-1)^r
{{M-\ell -1} \choose {r}}
(\ell+r+1)
E \left [
\phi^{r+\ell}(t)
\right ]
\ee
where $E[\cdot]$ is the expectation using $f_T(t)$.

\subsection{$H(\Omega|\Svv,\Tmat)$ Exponential First Passage}
\label{sect:specialcaseHO}
Here we derive an expression for $I(\Svv;\Tmat)$ when the input distribution is
that which maximizes $I(\Smat;\Tmat)$ subject to exponential first passage and
an emission deadline -- we assume $f_T(t)$ is limited to the interval
$[0,\tau]$.  The $f_T()$ that maximizes $h(S)$ was derived in \cite{isit11} as
\be
\label{eq:iidTdeadline}
f_{T_m}(t)
=
\frac{1}{e + \lambda \tau}
\delta(t)
+
\frac{\lambda}{e + \lambda \tau}
[u(t) - u(t - \tau)]
+
\frac{e-1}{e + \lambda \tau}
\delta(t-\tau)
\ee
$m=1,2,...,M$.

To obtain $\Hup(\Tmat)$ we calculate
\be
\label{eq:phiISalmostfs}
\phi(t)
=
\int_0^{t}
f_T(x) e^{-\lambda (t-x)} dx
=
\threedef{
{\frac{1}
{e + \lambda \tau}}}{0\le t \le \tau}
{\frac{e}{e + \lambda \tau}
e^{-\lambda (t - \lambda \tau)}}{t \ge \tau}{0}{\mbox{o.w.}}
\ee
We require an expression for the integral $\int_0^{\infty} f_T(t)
\phi^k(t)dt$. Since $\int_{0^-}^{0^+} \delta (t) u^k(t) dt =
\frac{1}{k+1}$ we obtain
$$
\begin{array}{rcl}
E_T \left [ \phi^k(T) \right ]
 & = & 
\left (\frac{1}{e+\lambda \tau} \right )^{k+1}
{\displaystyle \int_{0^-}^{0^+} }
\delta(t) 
u^k(t)
dt
+
\lambda\left (\frac{1}{e+\lambda \tau} \right )^{k+1}
\int_0^{\tau}
dt\\
 & + & 
(e-1)\left ( \frac{1}{e+\lambda \tau} \right )^{k+1}
{\displaystyle \int_{\tau^-}^{\tau^+} }
\delta(t-\tau) 
\left ( 1 + (e-1)u(t-\tau) \right )^k
dt
\end{array}
$$
which reduces to
$$
\left (\frac{1}{e+\lambda \tau} \right )^{k+1}
\left [
\frac{1}{k+1} + \lambda \tau
+
\sum_{r=0}^k
{k \choose r}
\frac{1}{r+1}
(e-1)^{r+1}
\right ]
$$
which further reduces to
$$
\left (\frac{1}{e+\lambda \tau} \right )^{k+1}
\left [
\frac{1}{k+1} +  \lambda \tau
+
\frac{e^{k+1}}{k+1}
-
\frac{1}{k+1}
\right]
$$
and then
$$
E_T \left [ \phi^k(T) \right ]
=
\left (\frac{1}{e+\lambda \tau} \right )^{k+1}
\left [
\lambda \tau
+
\frac{e^{k+1}}{k+1}
\right]
$$
so that \equat{Gammadiffgeneral} becomes
$$
\overline{\Delta\Gamma}_{M,\ell} 
=
{{M} \choose {\ell+1}}
\sum_{r=0}^{M-\ell -1}
(-1)^r
{{M-\ell -1} \choose {r}}
(\ell+r+1)
\left (\frac{1}{e+\lambda \tau} \right )^{r+\ell+1}
\left [
\lambda \tau
+
\frac{e^{r+\ell+1}}{r+\ell+1}
\right]
$$
which reduces to 
\begin{align*}
\overline{\Delta\Gamma}_{M,\ell}  &= {{M} \choose {\ell+1}}
\left (\frac{e}{e+\lambda \tau} \right )^{\ell+1}
\left (\frac{\lambda \tau}{e+\lambda \tau} \right )^{M-\ell-1}\\
&+
\lambda \tau
{{M} \choose {\ell+1}}
\sum_{r=0}^{M-\ell -1}
(-1)^r
{{M-\ell -1} \choose {r}}
(\ell+r+1)
\left (\frac{1}{e+\lambda \tau} \right )^{r+\ell+1}
\end{align*}
and then to
\begin{align*}
\overline{\Delta\Gamma}_{M,\ell} &= {{M} \choose {\ell+1}}
\left (\frac{e}{e+\lambda \tau} \right )^{\ell+1}
\left (\frac{\lambda \tau}{e+\lambda \tau} \right )^{M-\ell-1}\\
&+
\lambda \tau
{{M} \choose {\ell+1}}
\left (1 - \frac{1}{e+\lambda \tau} \right )^{M-\ell -2}
\left (\frac{1}{e+\lambda \tau} \right )^{\ell +1}
\left ( \ell + 1 -\frac{M}{e+\lambda \tau} \right )
\end{align*}
If we define $k = \ell +1$ and then
$$
p_1 = \frac{e}{e+\lambda \tau}
$$
and
$$
p_2 = \frac{1}{e+\lambda \tau}
$$
we can then write
n\be
\label{eq:Gammadiffuniform}
\overline{\Delta\Gamma}_{M,k-1}
=
{M \choose k}
p_1^k (1-p_1)^{M-k}
+
\frac{\lambda \tau }{1-p_2}
\left [
k
-
\frac{M}{\lambda \tau + e}
\right ]
{M \choose k}
p_2^k (1-p_2)^{M-k}
\ee
Now if we define random variables $K_i$ to be binomial over $M$ trials and
success probability $p_i$, then we have in theorem form:
\begin{theorem}
\thmlabel{HOmegadeadlineT}
For exponential first passage with parameter $\lambda$, a launch deadline constraint
$\tau$ and the corresponding $I(\Smat;\Tmat)$-maximizing launch density
\equat{iidTdeadline} we have
\be
\label{eq:HOexpgeomform2}
H_e(\Omega|\Svv,\Tmat)
=
E_{K_1} \left [
\log K_1!
\right ]
+
E_{K_2} \left [
\left ( K2\frac{\lambda \tau}{1-p_2}
-
\frac{\lambda \tau M }{(1-p_2)(\lambda \tau + e)} \right )
\log K_2!
\right ]
\ee
where $K_1$ and $K_2$ are a binomial random variables over $M$ trials with success probabilities
$p_1 = \frac{e}{e+\lambda \tau}$ and $p_2 = \frac{1}{e+\lambda \tau}$, respectively.
\end{theorem}

And since the associated maximized $I(\Smat;\Tmat)$ is $M \log (1 +
\frac{\lambda\tau}{e})$ \cite{isit11} we then have the following Lemma:
\begin{lemma}
\lemlabel{IdeadlineT}
For exponential first passage with parameter $\lambda$, a launch deadline of
$\tau$ and $f_{\Tmat}()$ given by \equat{iidTdeadline} we have $I(\Svv;\Tmat)$ as
\be
\label{eq:Igeomform2}
M \log (1 + \frac{\lambda\tau}{e})
-
\log M!
+
E_{K_1} \left [
\log K_1!
\right ]
+
E_{K_2} \left [
\left ( K2\frac{\lambda \tau}{1-p_2}
-
\frac{\lambda \tau M }{(1-p_2)(\lambda \tau + e)} \right )
\log K_2!
\right ]
\ee
where $K_1$ and $K_2$ are a binomial random variables over $M$ trials with success probabilities
$p_1 = \frac{e}{e+\lambda \tau}$ and $p_2 = \frac{1}{e+\lambda \tau}$, respectively.
\end{lemma}

\section{Lower Bounds on Channel Capacity}
In the introduction we defined a channel use as the launch and capture of $M$
quanta under mean and deadline constraints on emission times. We then assumed
sequential (or parallel) independent channel uses so that the figure of merit
was the mutual information $I(\Svv;\Tmat)$.  Here we use the results we have
derived to consider information flow limits under more physically plausible
conditions in systems where channel uses are not so crisply defined {\em a priori}.

For instance, energy is a key resource in biological systems.  Thus, a good
figure of merit for biological communication efficiency is nats/joule.  In the
current context, a natural definition of capacity would be nats/quantum since
signal molecule construction (often a protein in biological systems) requires a
known amount of energy.  At roughly $4$ ATP per amino acid \cite{lehninger2005},
construction of a $100$-amino acid protein would require $400$ ATP -- a
significant cost even in comparison to an elevated $6 \times 10^4$ ATP/sec total
energy budget during cell replication (E. Coli \cite{lehninger}) when one
considers that many signaling molecules must be produced.  Thus, it makes sense
to rewrite emission time constraints as a constraint on average quantum production
$\rho$ (quanta/second).  Our previous emission constraint is then
\be
\label{eq:rhodef}
\tau = \tau(M)
= \frac{M}{\rho}
\ee

\begin{wrapfigure}{r}{3.3in}
\vspace{-0.4in}
\begin{center}
\includegraphics[height=0.75in,width=3.0in]{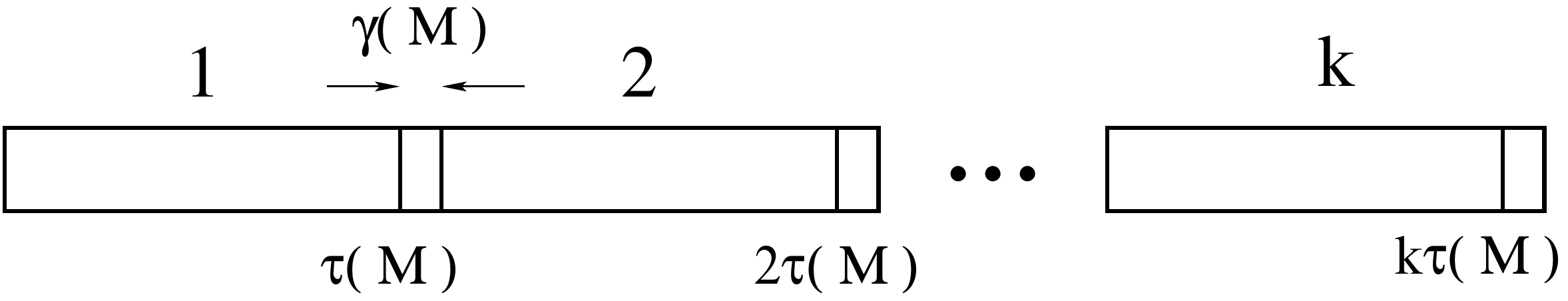}
\end{center}
\vspace{-0.25in}
\caption{\small Successive $M$-emission channel uses.}
\label{fig:channeluse}
\end{wrapfigure}

So, consider FIGURE~\ref{fig:channeluse} where sequential transmissions of $M$
quanta -- channel uses -- are depicted.  We will assume a ``guard interval'' of
some duration $\gamma(M,\epsilon)$ between successive transmissions 
so that all $M$ transmissions are received
before the beginning of the next channel use
with high probability ($1-\epsilon$).

We further require that the average emission rate, ${M}/(\tau(M) + \gamma(M,\epsilon))$ satisfies
\be
\label{eq:limitrho}
\lim_{\epsilon \rightarrow 0}
\lim_{M\rightarrow \infty}
\frac{M}{\tau(M) + \gamma(M,\epsilon)}
=
\rho
\ee

A convenient choice of  $\gamma(M,\epsilon)$ is $\epsilon \tau(M)$ for any $\epsilon > 0$.   We then require that
\be
\label{eq:timelimit}
\lim_{M \rightarrow \infty}
\mbox{Prob}\{\Svv_M \le \tau(M)(1 + \epsilon) \} = 1
\ee
We can interpret \equat{timelimit} as given arbitrarily small $\epsilon$
we can always find a finite $M^*$ such that
$$
\mbox{Prob}\{\svv_M \le \tau(M)(1 + \epsilon) \} > 1-\epsilon
$$
$\forall M  \ge  M^*$.
We can now derive conditions on first passage time densities under which
\equat{timelimit} is true.

Calculating a CDF for $\Svv_M$ is in general difficult since emission times
$T_m$ might be correlated. However, for a fixed emission interval
$[0, \tau(M)]$ we can readily calculate a worst case CDF for $\Svv_M$ and thence
a deterministic upper bound on the actual signaling epoch duration that is
satisfied with probability $1-\epsilon$.  That is,
for a given emission schedule $\tv$, the CDF for the final arrival is
$$
F_{\Svv_M|\tv}(s|\tv) = \prod_{m=1}^M  G(s-t_m)u(s-t_m)
$$
so that
$$
F_{\Svv_M}(s)  = \int_{\bzero}^{{\bf \tau(M)}} f_{\Tmat}(\tv)\prod_{m=1}^M  G (s-t_m)u(s-t_m) d \tv
$$
However, it is easy to see that
$$
F_{\Svv_M}(s) \ge G^M(s-\tau(M)) u(s-\tau(M))
$$
since $G(s-t_m)$ is monotone decreasing in $t_m$.

For $s = \tau(M)(1+\epsilon)$ we have
\be
\label{eq:timelimit2}
F_{\Svv_M}(\tau(M)(1+\epsilon))
\ge
G^M(\frac{M\epsilon}{\rho})
\ee
and we require
\be
\label{eq:timelimit2a}
\lim_{M \rightarrow \infty}
G^M(\frac{M\epsilon}{\rho})
=
1
\ee
which for convenience, we rewrite as
\be
\label{eq:timelimit3}
\lim_{M \rightarrow \infty}
M \log G(\frac{M\epsilon}{\rho})
=
0
\ee
Thus, to satisfy \equat{timelimit3}, $(\log G(\epsilon\frac{M}{\rho}))^{-1}$
must be asymptotically supralinear in $M$.

If rewrite $\log G(\frac{M\epsilon}{\rho})$ in terms of the CCDF $\bG()$ and note that
$\log(1-x) \approx -x$ for $x$ small, we have
$$
\bG(\frac{M\epsilon}{\rho}) - \epsilon
\le
\log \left ( 1 - \bG(\frac{M\epsilon}{\rho}) \right )
\le
\bG(\frac{M\epsilon}{\rho}) + \epsilon
$$
for sufficiently large $M$.  Thus, a first passage distribution whose CCDF satisfies
\be
\label{eq:timelimit4}
\lim_{M \rightarrow \infty}
M \bG(\frac{M\epsilon}{\rho})
=
0
\ee
will also allow satisfaction of \equat{timelimit} with $\tau(M) = \frac{M}{\rho}$ and $\gamma(M,\epsilon) = \epsilon \tau(M)$.

Since all first passage times are non-negative random variables, 
\be
\label{eq:meanintegral}
E[D]
=
\int_0^{\infty}
\bG(x) dx
\ee
The integral exists {\bf iff}  $1/\bG(x)$ is asymptotically supralinear in
$x$. Thus, if the mean first passage time $E[D]$ exists, then \equat{timelimit4}
is satisfied.  Finally, in the limit of vanishing $\epsilon$ we have
$$
\lim_{\epsilon \rightarrow 0}
\lim_{M \rightarrow \infty}
\frac{M}{\tau(M) + \gamma(M,\epsilon)}
=
\lim_{\epsilon \rightarrow 0}
\frac{\rho}{1+\epsilon}
=
\rho
$$
as required by \equat{limitrho}

\subsection{Capacity Lower Bound in Nats Per Quantum}
\label{sect:natsperquantum}
The maximum mutual information between $\Tmat$ and $\Svv$ per quantum given $M$
launched quanta with timing constraint $\tau(M) = M/\rho$ is
\be
\label{eq:perquantumCtauM}
C_q(M)
=
\frac{1}{M} 
\max_{f_{\Tmat}()} I(\Svv;\Tmat)
\ee
We define the limiting capacity in nats per quantum as
\be
\label{eq:perquantumCtauinf}
C_q
=
\lim_{M \rightarrow \infty}
C_q(M)
\ee
$C_q(M)$ will be monotone increasing in $M$ since concatenation of two
emission intervals with durations $\tau/2$ and $M/2$ quanta each is more
constrained than a single interval of duration $\tau$ with $M$ quanta.

We can derive a simple lower bound on $C_q(M)$ by noting that
\equat{orderedMI_decomp} and the definition of \equat{perquantumCtauM} with
$\tau(M)$ produces
\be
\label{eq:RtauMboundsimple}
C_q(M)
=
\max_{f_{\Tmat}()}
\left [  I(\Smat;\Tmat) + H(\Omega|\Svv,\Tmat) \right ]
-
\log M!
\ge
\max_{f_{\Tmat}()}
I(\Smat;\Tmat) - \log M!
\ee
because $0 \le H(\Omega|\Svv,\Tmat) \le M!$.  

From \cite{isit11} we know that the univariate maximum $I(S;T)$
subject to $T \le \tau$ and a mean first passage time $\lambda^{-1}$ is also
minimized when the mean first passage time density $g()$ is exponential with
parameter $\lambda$.  Therefore, via Lemma~\ref{lem:IdeadlineT} we have for any
finite $M$ and a finite launch deadline $\tau(M)$,
\be
\label{eq:Iminmaxdeadline}
I(\Smat;\Tmat)
\ge
\min_{g()} \max_{f_{\Tmat}()} I(\Smat;\Tmat)
=
M \log \left (1 + \frac{\lambda \tau(M)}{e} \right )
\ee
which means,
\be
C_q(M)
\ge
\log \left (1 + \frac{\lambda \tau(M)}{e} \right ) - \frac{\log(M!)}{M}
\ee
for a launch deadline $\tau(M)$.

Using \equat{rhodef} and Stirling's approximation, $\log M! = M \log M - M + O(\log(M))$ we have
$$
\begin{array}{rcl}
\frac{1}{M} \left ( M \log \left ( 1 +  \frac {\lambda}{\rho e} M \right ) - \log M! \right )
 & = &
\log \left ( 1 + \frac{\lambda}{\rho e} M \right )
- \log M + 1 - \frac{1}{M}O(\log(M))\\
 & = &
\log \left ( \frac{e}{M}  + \frac{\lambda}{\rho} \right )
- \frac{1}{M}O(\log(M))
\end{array}
$$
Defining $\chi = \frac{\lambda}{\rho}$, the ratio of the uptake rate to the
release rate, and then taking the limit as $M \rightarrow \infty$ we obtain
\be
\lim_{M \rightarrow \infty}
C_q(M)
=
\log \chi
\ee
We summarize the results with a theorem:
\begin{theorem}
  \thmlabel{IDlowerbound}

{\bf [$C_q$ Lower Bound with Emission Deadline $\tau(M) = \frac{\rho}{M}$ ]}

Given an average rate of signaling quantum production $\rho$ as defined in
\equat{limitrho} and any i.i.d. first passage time distribution with mean $\lambda
^{-1}$, the timing channel capacity $C_q(\chi)$ in nats per quantum obeys
\be
\label{eq:Cqlo}
C_{q}(\chi) \ge \max \left
    \{\log\chi,0 \right \}
\ee
where $\chi = \frac{\lambda}{\rho}$
\end{theorem}
We emphasize that the Theorem \thmref{IDlowerbound} bound is {\em general} and
applies to {\em any} first passage time density $g()$ with mean $\lambda^{-1}$.

\subsection{Capacity Lower Bound in Nats Per Unit Time}
\label{sect:natspertime}
The duration of a signaling epoch is $\tau(M) + \gamma(M,\epsilon)$.  Thus, for a given
number $M$ of emissions per channel use we define the channel capacity in nats per unit time as
$$
C_t(M)
=
\max_{f_{\Tmat}()} \frac{I(\Svv;\Tmat)}{\tau(M) + \gamma(M,\epsilon)}
=
C_q(M, \tau(M)) \frac{M}{\tau(M) + \gamma(M,\epsilon)}
$$
where the $C_q(M, \tau(M))$ explicitly denotes an emission interval of duration $\tau(M)$.  However, since we define
\be
\label{eq:limitrhoM}
\rho
=
\frac{M}{\tau(M) + \gamma(M,\epsilon)}
\ee
we then have
\be
\label{eq:Ctf1M}
C_t(M) = \rho C_q\left (M, M \left (\frac{1}{\rho} - \frac{\gamma(M,\epsilon)}{M} \right ) \right ) 
\ee

For any given tuple $(\rho, M, \epsilon)$, a positive interval duration
$\tau(M)$ such that all quanta are received by the end of the signaling epoch,
$\tau(M) + \gamma(M,\epsilon)$, either exists or does not.  So, assume that a
valid $\tau(M)$ exists.  We know from the previous section that $2C_q(M/2) \le
C_q(M)$.  We also know that
$$
C_q\left (M, \tau(M) - \alpha \right )
\le 
C_q\left (M, \tau(M)  \right )
$$
for $\alpha > 0$ since increasing the emission interval cannot decrease the
maximum mutual information.  We also know from the previous section that if $E[D]$ exists, then the
guard interval duration, $\gamma(M,\epsilon)$ can be sublinear in $M$.  So, if we set $\tau(M) =
M/\rho$, then $C_q\left (M, M \left (\frac{1}{\rho} -\frac{\gamma(M,\epsilon)}{M} \right ) \right )$
is an increasing function of $M$ whose limit is $C_q$.  We summarize with the 
following theorem:
\begin{theorem}
  \thmlabel{timelimit_firstpassage}
If $E[D]$ exists, then the capacity in nats per unit time of the {\bf quantum release timing channel} obeys
\be
\label{eq:CtisrhoCq}
C_t
=
\rho
C_q
\ee
where $C_q$ is defined in \equat{perquantumCtauinf} and $\rho$ is the average quantum emission rate.
\end{theorem}

\subsection{Special Case Lower Bounds: exponential first passage}
Given exponential first passage, Lemma~\ref{lem:IdeadlineT} provides a lower bound on $I(\Svv;\Tmat)$ 
for a deadline launch constraint.  We now examine
$$
\lim_{M \rightarrow \infty}
\frac{I(\Svv;\Tmat)}{M}
$$
where we assume the launch constraint is specified by $\tau(M) = \frac{M}{\rho}$ as
in sections~\ref{sect:natsperquantum} and \ref{sect:natspertime}.  To begin, remember that
$$
\lambda \tau(M) = \frac{\lambda}{\rho} M
\equiv
\chi M
$$
and then note that
$
{M \choose k} 
\left ( \frac{1}{1 + \chi M} \right )^k
\left (1 -  \frac{1}{1 + \chi M} \right )^{M-k}
$
\blankout{
$$
=
\frac{M!}{(M-k)!k!}
\left ( \frac{1}{\chi M} \right )^M
\left (\frac{1}{\frac{1}{M \chi} + 1} \right )^M
(M \chi)^M
\left ( \frac{1}{\chi M} \right )^k
$$
}
reduces to
$$
\frac{M(M-1) \cdots (M-k+1)}{M^k k!}
\left (\frac{1}{\frac{1}{M \chi} + 1} \right )^M
\left ( \frac{1}{\chi} \right )^k
$$
For any finite $k$ it is easily seen that
\be
\label{eq:limit1}
\lim_{M \rightarrow \infty}
{M \choose k} 
\left ( \frac{1}{1 + \chi M} \right )^k
\left (1 -  \frac{1}{1 + \chi M} \right )^{M-k}
=
e^{-\frac{1}{\chi}}
\frac{1}{k!}
\left ( \frac{1}{\chi} \right )^k
\ee
Similarly
\be
\label{eq:limit2}
\lim_{M \rightarrow \infty}
{M \choose k} 
\left ( \frac{1}{e + \chi M} \right )^k
\left (1 -  \frac{1}{e + \chi M} \right )^{M-k}
=
e^{-\frac{1}{\chi}}
\frac{1}{k!}
\left ( \frac{1}{\chi} \right )^k
\ee
and we also have
\be
\label{eq:limit3}
\lim_{M \rightarrow \infty}
{M \choose k} 
\left ( \frac{e}{e + \chi M} \right )^k
\left (1 -  \frac{e}{e + \chi M} \right )^{M-k}
=
e^{-\frac{e}{\chi}}
\frac{1}{k!}
\left ( \frac{e}{\chi} \right )^k
\ee

\Equat{HOexpgeomform2} can be combined with
\equat{limit1}, \equat{limit2} and \equat{limit3} to produce the following theorem:
\begin{theorem}
\thmlabel{C2}
{\bf [Exponential First Passage $C_q$ lower bound: emission deadline]}

For exponential first passage and $\Tmat \in [0,M/\rho]^M$, the channel capacity
in nats per quantum obeys
\be
\label{eq:C2}
C_{q}(\chi)
\ge
\log \chi
+
e^{-\frac{1}{\chi}}
\sum_{k=2}^{\infty}
\left (\frac{1}{\chi} \right )^k
(k \chi - 1)
\frac{\log k!}{k!}
\ee
\end{theorem}

\blankout{
A moderately tight lower bound can be derived for \equat{C2} since 
$\left (\frac{1}{\chi} \right )^k \chi (k \chi - 1)\frac{1}{k!}$ is a PMF ($k \ge 2$) with mean
$1 + (e^{\frac{1}{\chi}}-1)\chi$ so long as $2\chi - 1 > 0$.
We then have via Jensen's inequality,
\begin{lemma}
\lemlabel{cibox2}
If $\chi \ge \frac{1}{2}$ then
\be
\label{eq:cibox2}
C_{q}(\chi)
\ge
\log \chi
+
\frac{e^{-\frac{1}{\chi}}}{\chi}
\log \left (
\Gamma \left (
2 + (e^{\frac{1}{\chi}}-1)\chi
\right )
\right )
\ee
where we define $\chi \equiv \frac{\lambda}{\rho}$ with
$\lambda$ the first passage rate and $\rho$ the quantum production rate.
\end{lemma}

We also note that the leftmost range of the bound in Lemma~\ref{lem:cibox2} can
be extended to $1/\ell$ by separating the sum in \equat{C2} into two parts ($k
\ge \ell$ and $2 \le k < \ell$) and applying Jensen's inequality to the upper
sum after appropriate normalizations to assure $\left (\frac{1}{\chi} \right )^k
\frac{(k\chi - 1)}{k!}$ is a probability mass function for $k \ge \ell$.
Although this procedure is not analytically informative, it does simplify
numerical computation of the bound.
}

\section{Discussion \& Conclusion}
We have described a basic model for a quanta timing channel wherein identical
quanta are released and travel independently to a receiver with
information conveyed by the timing of arrivals.  We have derived general
machinery for the analysis of such channels and provided lower bounds on channel
capacity under the assumption that the mean first passage time between sender
and receiver is finite.  The lower bounds on capacity are on the order of a half
nat per first passage time.

\begin{wrapfigure}{r}{3.0in}
\vspace{-0.4in}
\begin{center}
\includegraphics[height=2.5in,width=3.0in]{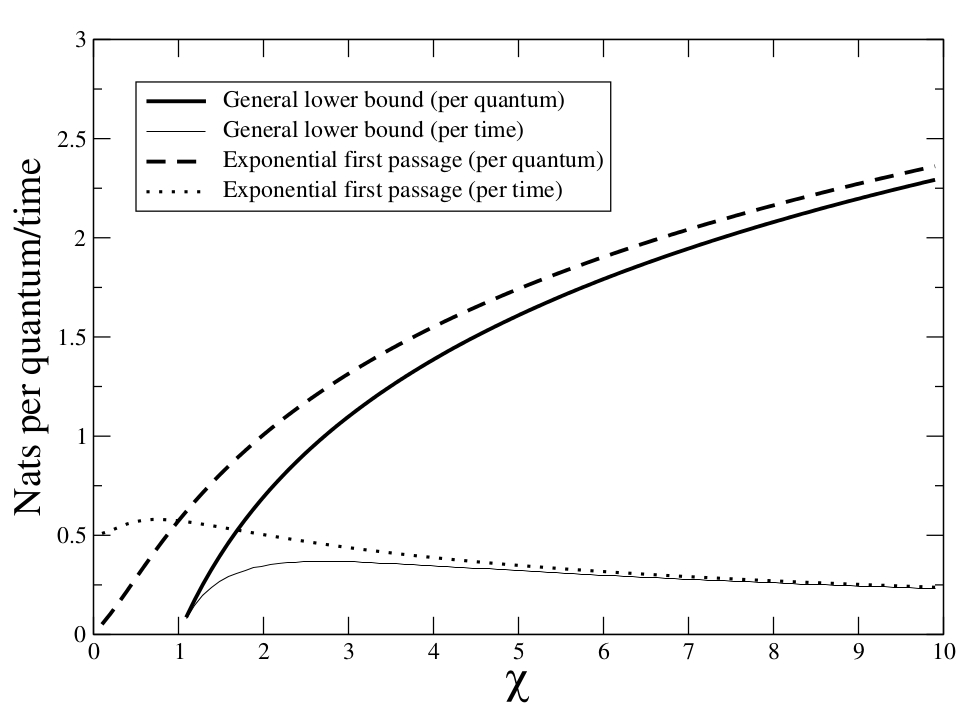}
\end{center}
\vspace{-0.3in}
\caption{\footnotesize Lower bounds for $C_q$ and $C_t$ vs. $\chi$.}
\label{fig:CqtElo}
\end{wrapfigure}
It is worth noting that free diffusion (brownian motion) first passage times are
{\em \bf not} finite and thus not well-behaved from an information theoretic
capacity standpoint.  However, in an finite spatial-extent system, physical
constraints on quanta motion enforce finite first passage.  It is also
noteworthy that by considering quanta in the limit of large $M$ per signaling
interval, our results in principle bridge the gap between quantum channel
descriptions and signaling agent concentration-based descriptions.  That is,
though the signaling problem formulation is epoch-based ($M$ quanta per emission
period $\tau$, and $\rho = M/\tau$ constant), with large $M$ (and concomitantly
large $\tau$), the ``instantaneous'' concentrations of quanta within an emission
period are not so constrained.

The question of quanta number vs. timing information is worth exploring briefly.
Consider that instead of fixing the number of quanta per epoch, we might send
different numbers of quanta in each epoch.  We have shown that $C_q(M)$ is at
least linear in $M$.  In contrast, the maximum amount of information conveyed
per epoch by the {\em number} of quanta is exactly $\log M$ -- strongly {\em
  sub}-linear in $M$.  The argument also applies to $C_t(M)$ since the guard
interval is proportionately larger for small $M$ (larger-$M$ intervals are more
temporally efficient and therefore higher rate).  Thus, in terms of information
transfer, timing information seems strongly preferred, at least asymptotically.

Our lower bounds on channel capacity in
nats per quantum (\equat{Cqlo} and \equat{C2}) and the corresponding bounds in
nats per unit time (\equat{CtisrhoCq}) are shown FIGURE~\ref{fig:CqtElo}.
Increasing $\chi$ increases the emission interval relative the mean first
passage time and thereby increases the information content of any individual
quantum.  In addition, since successive quanta may be less likely to interchange
position, $\frac{1}{M} \left ( \log M! - H(\Omega|\Svv;\Tmat) \right )$
approaches zero.  Thus, the simple lower bound of \equat{Cqlo} (and
correspondingly \equat{CtisrhoCq}) meets the lower bound for exponential first
passage which has minmax $I(\Smat;\Tmat)$. But perhaps most interesting is the
implication that there may exist optimum emission rates for a given channel as
evidenced by the shape of the $C_t$ curves in FIGURE~\ref{fig:CqtElo}. This
feature echos \cite{fekriisit11} where an optimum burst interval for signaling
molecules in a diffusive channel was derived. However, since we do not know the
channel capacity, we do not know how tight our lower bounds are.  It is
therefore premature to say whether an optimum emission rate is a feature of the
identical quanta timing channel.

\vspace{-0.4in}
\begin{singlespace}
\scriptsize
\bibliography{MERGE11,sm_collection}
\bibliographystyle{unsrt}
\end{singlespace}
\end{document}